\documentstyle[twoside,fleqn,espcrc2,epsf]{article}

\newcommand{\AmS}{{\protect\the\textfont2
  A\kern-.1667em\lower.5ex\hbox{M}\kern-.125emS}}
\newcommand{\Tr}{{\rm Tr}}

\def\lsim{\raise0.3ex\hbox{$<$\kern-0.75em\raise-1.1ex\hbox{$\sim$}}}
\def\gsim{\raise0.3ex\hbox{$>$\kern-0.75em\raise-1.1ex\hbox{$\sim$}}}

\title{
\vspace{-4.0cm}
\begin{flushright}
{\normalsize
ITP-Budapest 546\\
UTCCP-P-46\\
UTHEP-389\\
\vspace{-.2cm}
September 1998}
\end{flushright}
\vspace{1.5cm}
The end point of the first-order phase transition
of the SU(2) gauge-Higgs model
on a four-dimensional isotropic lattice   
\thanks{Talk presented by Y.~Aoki at Lattice 98, Boulder, USA.}}

\author{
  Y.~Aoki\address{Center for Computational Physics, University of Tsukuba,
    Ibaraki 305-8577, Japan}$^,$\address{Institute of Physics,
    University of Tsukuba, Ibaraki 305-8571, Japan},
  F.~Csikor\address{Institute for Theoretical Physics,
    E\"otv\"os University, H-1088 Budapest, Hungary},
  Z.~Fodor$^{\rm c}$, and
  A.~Ukawa$^{\rm a , b}$
}

\begin{document}

\begin{abstract}
We report results of a study of the end point of the electroweak phase 
transition of the SU(2) gauge-Higgs model defined on a four-dimensional 
isotropic lattice with $N_t=2$.
Finite-size scaling study of Lee-Yang zeros yields $\lambda_c=0.00116(16)$ for 
the end point.  Combined with a zero-temperature measurement of Higgs 
and $W$ boson masses, this 
leads to $M_{H,c}=68.2\pm 6.6$ GeV for the critical Higgs boson mass. 
An independent analysis of Binder cumulant gives a consistent value 
$\lambda_c=0.00102(3)$ for the end point. 
\end{abstract}

\maketitle

\section{Introduction}

Recently study of the end point of the first-order 
electroweak phase transition has been pursued \cite{Laine98}
since it provides a basic information on the feasibility of electroweak 
baryogenesis. 
For the SU(2) gauge-Higgs model, detailed results 
have been reported within the three-dimensional effective theory 
approach \cite{Kjantie,KarschLat96,Lipzig97,Rumm98} 
including estimates of the critical Higgs boson mass at the end 
point \cite{KarschLat96,Lipzig97,Rumm98}.
Four-dimensional simulations, employing 
space-time anisotropic lattices \cite{Csikor981}
to overcome a multi-scale problem 
near the end point, have also been carried out \cite{Csikor982}.

In this article we report on results of our study of the end point using 
four-dimensional space-time symmetric lattices, building upon a 
previous work \cite{yaoki}. While computationally demanding, this 
approach is conceptually straightforward, whose results provide 
an independent check on those obtained with other methods.  

\section{Finite-Temperature Simulation}

We work with the standard SU(2) gauge-Higgs model Lagrangian given by 
\begin{equation}
\label{eq:S_lat}
  S =  S_g
    \mbox{} + \mathop{\sum}_x \{ \mathop{\sum}_\mu 2 \kappa L_{x, \mu}
    \mbox{} - \rho_x^2 - \lambda ( \rho_x^2 -1 )^2
    \},
\end{equation}
\begin{equation}
  \label{eq:link}
    L_{x, \mu}\equiv \frac{1}{2}
    \Tr ( \Phi_x^{\dag} U_{x, \mu} \Phi_{x+\hat{\mu}} ),
    \  \rho_x^2\equiv\frac{1}{2}\Tr (\Phi_x^{\dag}\Phi_x),
\end{equation}
where $S_g$ is the plaquette gauge action with the gauge 
coupling $\beta=4/g^2$. 
This work is made for the temporal extension $N_t=2$. 
For spatial lattices we employ $N_s^3=20^3,24^3,32^3,40^3,50^3$ and $60^3$.
We set the gauge coupling $\beta = 8$, and make
simulations at five  values of the scalar self-coupling $\lambda=0.00075,
0.001, 0.00135, 0.00145$ and $0.0017235$, which cover the range of Higgs
boson mass given by $57 \lsim M_H \lsim 85$GeV \cite{yaoki}.
The Higgs field hopping parameter $\kappa$ is tuned to the vicinity of 
the pseudo critical point. For each parameter point 
at least $10^5$ iterations of hybrid overrelaxation are made.
For further technical details we refer to Ref.~\cite{yaoki}.

\section{Analysis of Lee-Yang Zeros}

\begin{figure}[t]
  \vspace{-18pt}
  \epsfxsize=7.6cm \epsfbox{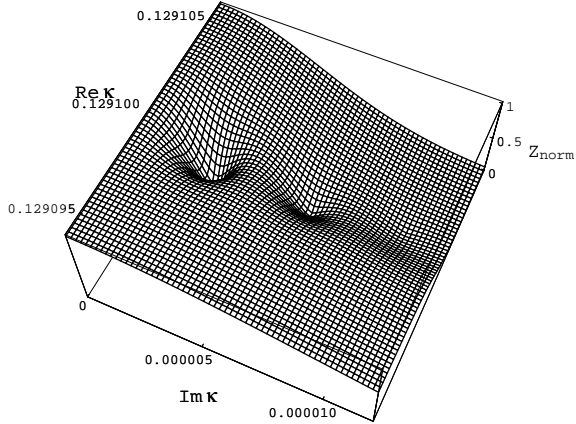}
  \vspace{-42pt}
  \caption{Absolute value of normalized partition function as a function
    of complex $\kappa$ for $\lambda=0.00075$ and $N_s=60$.}
  \label{fig:z_abs}
  \vspace{-12pt}
\end{figure}

We consider Lee-Yang zeros of the partition function
on the complex $\kappa$-plane at fixed $\lambda$.
To search for the zeros we carry out reweighting in $\kappa$ in both 
imaginary and real directions.
In Fig.~\ref{fig:z_abs} we show the absolute value
of the partition function normalized by its value at the real axis,
\begin{equation}
  \label{eq:z_norm}
  Z_{norm} \equiv
  \left| Z({\rm Re}\kappa,{\rm Im}\kappa) / Z({\rm Re}\kappa,0) \right|
\end{equation}
measured at $\lambda=0.00075$ and $N_s=60$.
We observe three zeros in this case. 

Let us call the zero nearest the real axis as first zero, and denote 
its location by $\kappa_0$. 
The value of $\kappa_0$ depends on spatial volume $V=N_s^3$ as well as 
on $\lambda$. 
The first-order phase transition of the model terminates when
the infinite volume limit $\kappa_0(V\to\infty)$ deviates away from the 
real axis. 

\begin{figure}[t]
  \hspace{-16pt}
  \epsfxsize=7.8cm \epsfbox{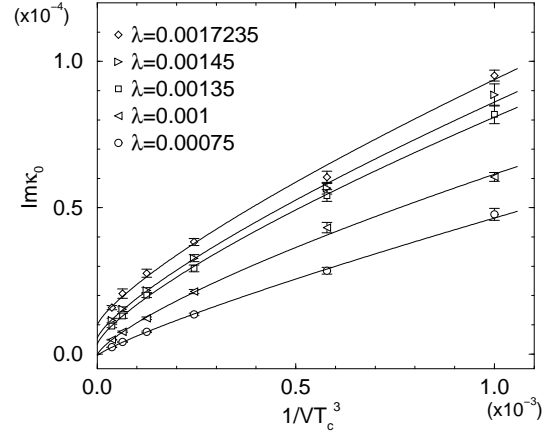}
  \vspace{-30pt}
  \caption{Imaginary part of first Lee-Yang zero as a function of inverse
    volume normalized by the critical temperature.
    Solid lines are least $\chi^2$ fits with 
    ${\rm Im}\kappa_0(V) = \kappa_0^c + C V^{-\nu}$.}
  \label{fig:fit_ly0}
  \vspace{-12pt}
\end{figure}

We show in Fig.~\ref{fig:fit_ly0} all of our results for the imaginary 
part of the first zero ${\rm Im}\kappa_0(V)$ as a function of volume. 
The infinite volume values are extracted by a $\chi^2$ fitting 
employing a scaling ansatz 
${\rm Im}\kappa_0(V) = \kappa_0^c + C V^{-\nu}$. 

\begin{figure}[t]
  \hspace{-12pt}
  \epsfxsize=7.8cm \epsfbox{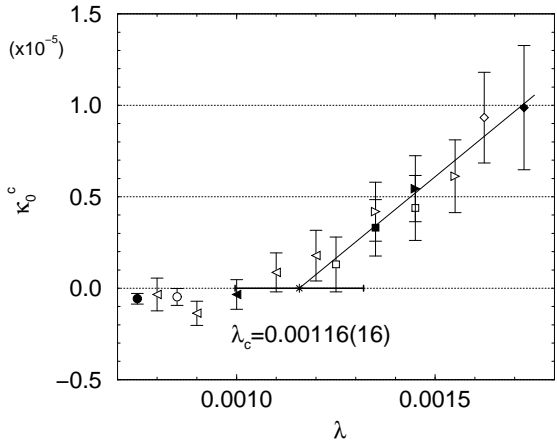}
  \vspace{-30pt}
  \caption{Imaginary part of first Lee-Yang zero at infinite-volume limit
    as a function of Higgs self coupling. Solid line is a linear fit to
    $\lambda=0.00135, 0.00145$ and $0.0017235$ (filled symbols).}
  \label{fig:res_ly0}
  \vspace{-15pt}
\end{figure}

Fit results for $\kappa_0^c$ are shown with filled symbols
in Fig.~\ref{fig:res_ly0}.
Open symbols are obtained by reweighting the partition function in 
the variable $\lambda$, starting from histogram results corresponding to 
the filled symbol of the same shape, 
and recalculating the first zero.  
The agreement of open symbols of different shapes within errors 
shows that reweighting from different values of $\lambda$ gives 
consistent results in the middle.

At small couplings $\lambda \lsim 0.001$, $\kappa_0^c$ is consistent
with zero, which agrees with the result of Ref.~\cite{yaoki}
that the transition is of first order in this region.
At large couplings $\lambda \gsim 0.0013$, $\kappa_0^c$ no longer
vanishes, and hence there is no phase transition.
In order to determine the end point of the phase transition, 
we take the three filled points at $\lambda=0.00135, 0.00145$ and $0.0017235$
directly obtained from independent simulations without $\lambda$-reweighting, 
and make a fit with a function linear in $\lambda$.
This gives the position of the end point to be $\lambda_c = 0.00116(16)$.

\section{Binder Cumulant}

\begin{figure}[t]
  \hspace{-12pt}
  \epsfxsize=7.8cm \epsfbox{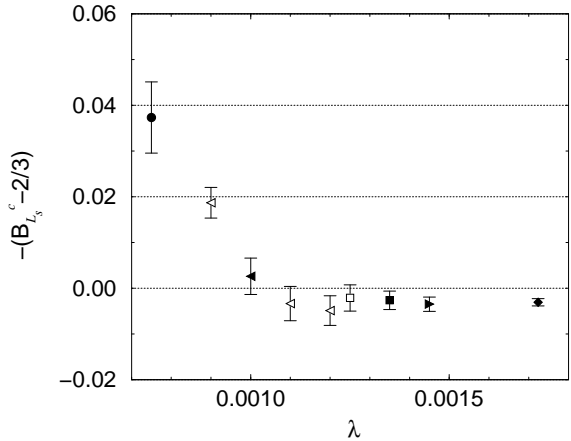}
  \vspace{-30pt}
  \caption{Minimum value of Binder cumulant of $L_s$
    at infinite volume limit as a function of $\lambda$.}
  \label{fig:bl}
  \vspace{-15pt}
\end{figure}

The Binder cumulant we study is defined in terms of 
$L_s=\sum_{x,\mu=1,2,3}L_{x,\mu}/3VN_t$ through
\begin{equation}
  \label{eq:binder}
  B_{L_s}(\kappa) \equiv 1
  \mbox{} - {\langle L_s^4 \rangle}/3{\langle L_s^2 \rangle^2}.
\end{equation}
We evaluate the minimum of the cumulant as a function of $\kappa$ 
for a given $\lambda$ and volume. 
We then use a scaling ansatz,
$B_{L_s}^{\rm min} = B_{L_s}^c + C V^{-\nu}$,
to extract the infinite-volume value.

In Fig.~\ref{fig:bl} we show $-(B_{L_s}^c - 2/3)$
as a function of $\lambda$, where meaning of symbols are the same as 
in Fig.~\ref{fig:res_ly0}.
A change of behavior from non-vanishing values to those consistent with zero 
at $\lambda \approx 0.001$ shows that the first-order phase transition 
terminates around this value. 
Linearly extrapolating the two independent data at $\lambda = 0.00075$
and $0.001$ yields $\lambda_c = 0.00102(3)$ for the end point, 
which is consistent with the result from our study of Lee-Yang zeros. 

\section{Critical Higgs Boson Mass}

To estimate the critical mass $M_{H,c}$ of the Higgs boson at the end 
point, we carry out a zero-temperature simulation
following the method of Ref.~\cite{Csikor96}.
Mass measurements are made at two points given by 
$(\lambda,\kappa=\kappa_c(\lambda,N_t=2))$ for $\lambda=0.0011$ and $0.00125$
employing several lattice sizes to control finite-volume effect.
Our preliminary results are 
$M_H=65.8\pm 0.6$ GeV ($\lambda=0.0011$) and 
$M_H=72.0\pm 0.4$ GeV $(\lambda=0.00125)$. 
Interpolating to the critical 
value $\lambda_c=0.00116(16)$ from the Lee-Yang zero analysis, 
we find $M_{H,c}=68.2\pm 6.6$GeV using $M_W=80$GeV as input. The error is 
dominated by that of $\lambda_c$.

\section{Conclusions}

Our result for the critical Higgs boson mass for $N_t=2$ is 
in agreement with the value $M_{H,c}=74.6\pm0.9$ GeV \cite{Csikor982} 
obtained in a four-dimensional anisotropic lattice simulation for the 
same temporal size.
The same work also reported that the critical mass decreases for larger 
temporal size, and extrapolates to $M_{H,c}=66.5\pm1.4$ GeV in the continuum
limit. This value is consistent with the three-dimensional 
result $67.0\pm0.8$ GeV \cite{Lipzig97}.  Thus results from various methods
converge well, and indicate that electroweak baryogenesis is unlikely within 
the Minimal Standard Model.
\vspace*{6pt}

Part of this work was carried out while ZF was visiting KEK by the Foreign 
Researcher Program of the Ministry of Education. This work is supported 
in part by Grants-in-Aid of the Ministry of Education (No. 10640246).

\end{document}